\documentclass[twocolumn,amsmath,amsfonts,amssymb,graphicx]{revtex4}

\newcommand{\z}{\par\addtocounter{items}{1}{\bf \arabic{items}.} }

\usepackage{graphicx}
\usepackage{epstopdf}

\begin{document}

\title{The Loewner equation: maps and shapes}

\author{Ilya A. Gruzberg and Leo P. Kadanoff}

\affiliation{The James Franck Institute, The University of Chicago\\
5640 S. Ellis Avenue, Chicago, Il 60637 USA\\
e-mail: gruzberg@uchicago.edu  and l-kadanoff@uchicago.edu}

\date{September 10, 2003} 

\begin{abstract}

\end{abstract}

\pacs{PACS numbers: }

\maketitle

The most exciting concepts in theoretical physics are those that
relate algebraic properties to geometrical ones. Of course, the
outstanding example of this is the geometric meaning of the
equations of general relativity, and their realization in the
shapes of possible universes and in black holes. Other examples
abound including the patterns of the paths of Brownian motion, the
forms of percolating clusters, the shapes of snowflakes and phase
boundaries, and the beautiful fingers of interpenetrating fluids
and of dendrites.

Especially beautiful and important patterns are seen in the case
in which these structures can be very large and fractal, as in a
long random walk or in the critical phenomena which occur near
higher order phase transitions. More than a century of effort has
gone into studying the problems posed by these objects. Much of it
has been devoted to figures in two dimensions. These shapes are
varied enough to be interesting, but relatively easy to
characterize and study.

In the last three years, new insights have permitted unexpected
progress in the study of fractal shapes in two dimensions.   A new
approach has arisen through analytic function theory and
probability theory, and given a new way of calculating fractal
shapes in critical phenomena, and in other problems like diffusion
limited aggregation (DLA), the theory of random walks, and of
percolation.

\section{Conformal Maps}

It all starts with the Riemann mapping theorem which gives a
method for characterizing shapes by using the theory of analytic
functions of a complex variable. High school students know how to
characterize a point $(x,y)$ in two-dimensional space by a complex
number $z=x+iy$.  With a little more thought they can understand
how a region $ \mathcal{D} $ in the $z$-plane that region might be
mapped into another region $ \mathcal{R}$ of the $w$-plane by the
function $w=g(z)$.

More advanced analysis considers the case in which $g$ is a
function which is analytic and univalent within $\mathcal{D}$. The
last property means that different points in $\mathcal{D}$ have
different images under the action of $g$ (or that $g$ does not
``glue'' points together). In this case it follows that the
derivative of $g(z)$ does not vanish within the region
$\mathcal{D}$. Then the mapping is called \mbox{ \it conformal},
and it takes the boundary of $\mathcal{D}$ into the boundary of
$\mathcal{R}$. Thus an analysis of functions of complex variables
automatically connects to a theory of the shapes of regions and
curves in two dimensions.

The Riemann mapping theorem states that any simply connected
region whose boundary has more then one point can be transformed
in an essentially unique manner into any other such region by a
conformal transformation.  We shall use the theorem to convert a
region of interest, for example a region contained in the upper
half plane, into a reference region, for example the entire upper
half plane, conventionally written as $\mathcal{H}$. To get all
the theorems we want, we shall need a few more constraints: our
regions are simply connected and contain the point at infinity.
The mappings take the form
\begin{align}
g(z) \to z + \frac{a}{z} + O(z^{-2}) \label{eq:asymptotic}
\end{align}
as $z$ goes to infinity.

\begin{figure}[]
\includegraphics[width=3in]{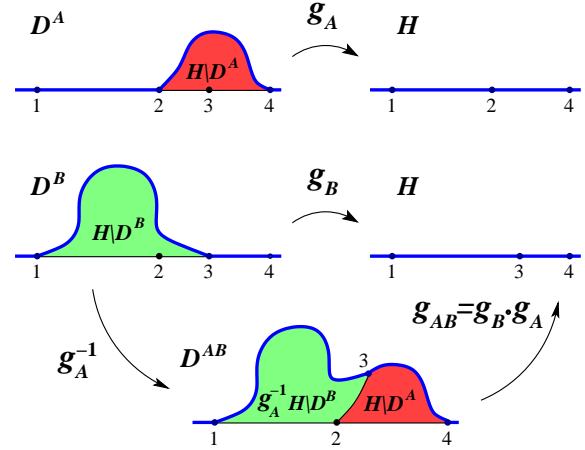}
\caption{Illustration for composition of conformal maps described
in the text. The notation $\mathcal{H}\backslash \mathcal{D}^A$
means the complement of $\mathcal{D}^A$ in $\mathcal{H}$.
Correspondence of points on the real axis and the boundary of
$\mathcal{D}^{AB}$ is shown.} \label{maps}
\end{figure}

The conformal mapping functions have a very important composition
property. Let the functions $g_A$ and $g_B$ respectively map the
regions $\mathcal{D}^A$ and $\mathcal{D}^B$ into the upper half
plane.  Then the composition of the two $g_{AB}(z)= g_B(g_A(z))$
maps a region $\mathcal{D}^{AB}$ which is contained within
$\mathcal{D}^A$ into the upper half plane (in the set-theoretic
notation $\mathcal{D}^{AB} = \mathcal{D}^A \backslash
g_A^{-1}(\mathcal{H}\backslash \mathcal{D}^B)$, see
Fig. \ref{maps} for an illustration). 
Another composition, giving $g=g_C \cdot g_B \cdot g_A $ maps a
subset of $\mathcal{D}^{AB}$ into the reference region. Successive
iterations gives us successively ``smaller" regions.

\section{Loewner equation}

To study this situation, Loewner defined families of maps, $g_t$,
defined by a real parameter $t$, which we might call time, and by
a real ``driving'' function $\xi(t)$. The Loewner equation
\begin{equation}
\frac{d}{dt} g_t(z) = \frac{2}{ g_t(z)-\xi(t)},  \quad \quad
g_{t=0}(z) = z \label{L}
   \end{equation}
defines a family of conformal maps $w=g_t(z)$ which take a
subregion of the upper half $z$-plane into the upper half
$w$-plane. The equation (\ref{L}) is written for a somewhat
arbitrary (but convenient) choice of the time variable, such that
the map $g_t(z)$ satisfies the asymptotic condition
(\ref{eq:asymptotic}) with $a = 2t$.

At time zero, $g_{t=0}$ is the identity map which maps
$\mathcal{H}$ into itself. The boundary curve is the real axis and
that too maps into itself. At each subsequent time $t$, $ g_t(z)$
defines a new mapping-region $\mathcal{D}_t$ which maps into
$\mathcal{H}$. Equally, the boundary of this region maps into the
real axis of the $w$-plane.

\begin{figure}[t]
\includegraphics[width=3in]{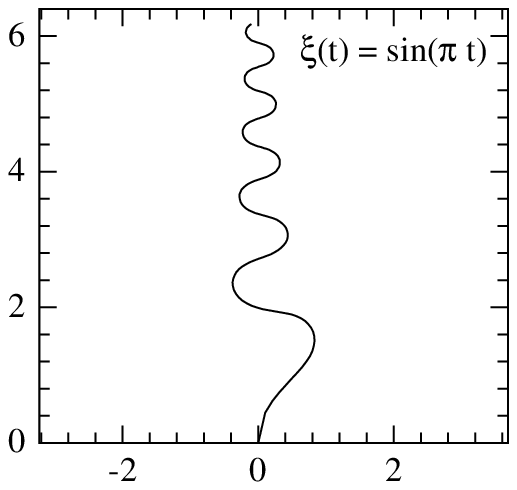}
\includegraphics[width=3in]{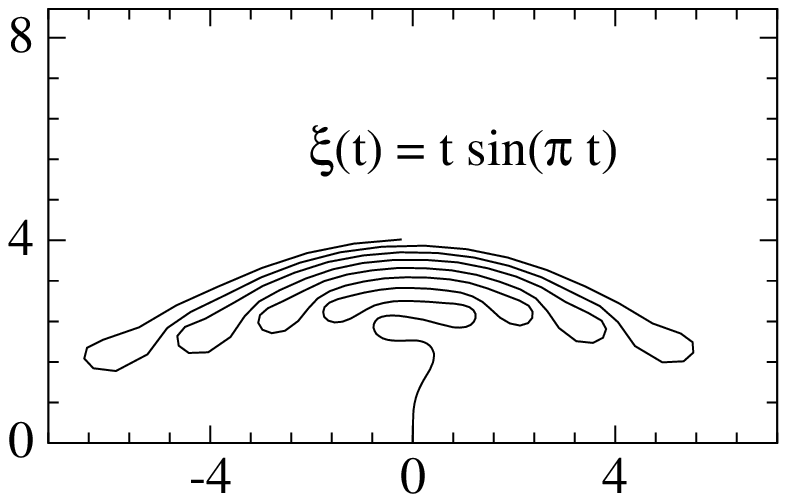}
\caption{The upper figure shows the trace for a periodic driving
function $\xi(t)= \sin(t)$. The trace looks similar at different
scales. The trace on the lower figure is driven by $\xi(t)= t
\sin(t)$.} \label{sin}
\end{figure}

The most important properties of $ g_t(z)$ arise from the
composition properties of conformal maps. Consider two maps
$g^A_t(z)$ and $g^B_t(z)$ generated respectively by the forcing
functions $\xi^A(t)$  and $\xi^B(t)$ which exist in the
respectively exist in intervals  $[0,t_A]$ and  $[0,t_B]$.  Now,
write $g_A=g_{t_A}^A(z) $ and $g_B=g_{t_B}^B(z) $   for the maps
generated by Eq. (\ref{L}) for each forcing and consider the
composite forcing:
\begin{eqnarray}
\xi(t) &= &  \xi^A(t)   \text{ for } (0 < t < t_A)  \nonumber  \\
  &= &\xi^B(t - t_A)   \text{ for } (t_A < t < t_A + t_B).
\end{eqnarray}
Then the composition rule indicates that the composite forcing
generates a map $g_t$ such that by the time $t_A + t_B$ it is
simply
\begin{equation} g_{t_A+ t_B}(z)= g_B(g_A(z)).
\label{composition}
\end{equation}
Indeed, fix $z$ and calculate $g_t(z)$ for $0 < t \leqslant t_A$
from Eq. (\ref{L}) using the initial condition $g_{t=0}(z) = z$.
Then repeat the calculation for $t_A < t \leqslant t_B$ with the
initial condition $g_{t=t_A}(z) = g_A(z)$. The result at $t=t_B$
is precisely described by the composition in the Eq.
(\ref{composition}).

An immediate consequence of this composition rule is that the
mapping sets $\mathcal{D}_t$ continually get ``smaller" as $t$
gets larger.  More properly stated if $s$ is greater than $t$ then
$\mathcal{D}_t$ contains $\mathcal{D}_s$.

In some sense, we can watch the mapping region get smaller. This
happens at time $t$ when some points $z_c(t)$ pass out of the
domain of analyticity of $g_t$.  From the Loewner equation that
will happen as the denominator in equation (\ref{L}) passes
through zero or at the points which obey
\begin{equation}
g_t(z_c(t)) = \xi(t). \label{non-A}
\end{equation}
Since $\xi(t)$ is real, that is, always on the boundary of $\cal
H$, its pre-image $z_c(t)$ always sits on the edge of the region
$\mathcal{D}_t$ that is mapped to $\cal H$ by the function
$g_t(z)$. If $\xi(t)$ is continuous, as time goes on the
singularities trace out a continuous curve in $\mathcal{H}$, which
we call simply the {\it trace}. The composition property implies
that if a point is in the trace now, it will remain in it for all
subsequent times. The trace is then a permanent path which shows
where the singularities arise and what points have been removed
from the mapping region.

\begin{figure}[t]
\includegraphics[width=6cm]{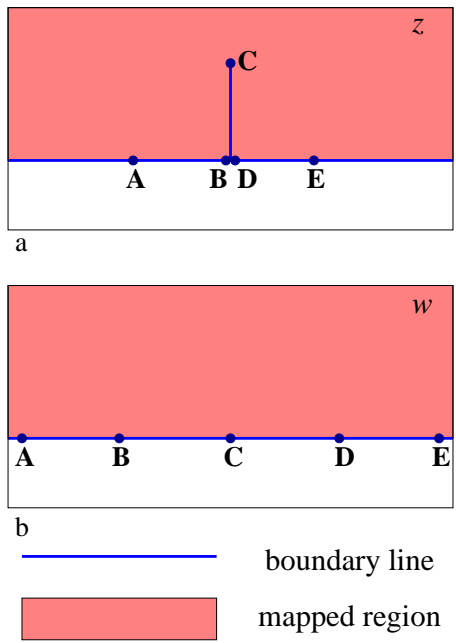}
\caption{This figure shows how a slit in the upper half of the
$z$-plane (part a) maps into a linear boundary in the $w$-plane
(part b). Notice that two neighboring  points in the $z$-plane
have images which are far apart in $w$. } \label{LEtraces1}
\end{figure}

The properties of the bounding curve composed of the trace and the
real axis, are absolutely amazing:
\begin{itemize}

\item if $\xi(t)$ is smooth enough so that its derivative exists
everywhere, the bounding curve never intersects itself.

\item if $\xi(t)$ is periodic,  the bounding curve is a
self-similar object.  See Figure \ref{sin} for an example. \item
the curve can intersect itself at finite time only if $\xi(t)$ is
sufficiently singular. The condition required
\cite{reftoAmsterdam} is that for some $t$
\begin{equation}
\lim_{\tau \rightarrow 0^+} \left| \frac{\xi(t
-\tau)-\xi(t)}{{\tau}^{0.5}}\right|
\end{equation}
goes to  a value greater than 4.
\end{itemize}
Thus the topological properties of the generated curve are
directly, but non-trivially,  related to the analytic properties
of the forcing function $\xi(t)$.

To see what happens in a specific example choose $\xi(t)$ to be a
constant, say $c$,  independent of time.  Then equation (\ref{L})
has the solution:
\begin{equation}
g_t(z) = c + \bigl((z-c)^2 + 4 t\bigr)^{1/2}
\end{equation}
As we can see in Figure \ref{LEtraces1} the singularity at time
$t$ is at $z_c(t)= c + 2 i  t^{1/2}$.  Thus, the trace is a
straight line which extends  from $z = c$ to $z = c + 2 i
t^{1/2}$.

\begin{figure}[t]
\includegraphics[width=6cm]{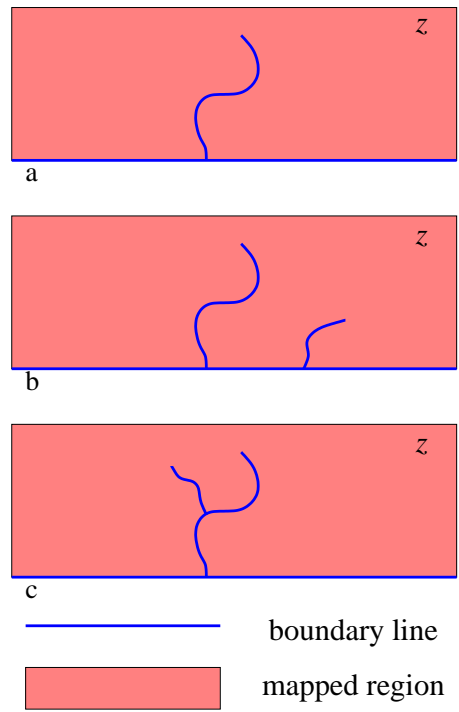}
\caption{Part a is a smooth curve produced by a differentiable
forcing function. Parts b and c show different traces produced by
a discontinuous $\xi(t)$.} \label{LEtraces2}
\end{figure}

More complex generating functions give much more complex boundary
curves. For example, imagine that at some time, $\xi(t)$ has a
discontinuous jump. The boundary then gains a new segment, see
Figure \ref{LEtraces2}, either having the new curve coming out of
the horizontal base or having it branch from the old segment. On
the other hand, if $\xi(t)$ varies sufficiently smoothly, the
boundary curve retains the topology shown in Figure
\ref{LEtraces2}a.  It is a smooth curve which is extended further
and further as time progresses, but never crosses itself. A few
exact solutions of the Loewner equation were obtained in a recent
paper \cite{NKK}.

\section{From Loewner to Critical Phenomena}

The years following Loewner's initial work showed many
applications of complex analytic methods to the study of problems
involving fractal or quasi-fractal objects in the plane.

One major approach was reformulation by Hastings and Levitov
\cite{Hastings-Levitov} of the DLA model originally put forward by
Witten and Sander \cite{Witten-Sander}. The latter authors
considered a step by step process in which an aggregate composed
of many soot particles grew into a large, fractal object. A new
tiny piece of soot would appear far away from the aggregate and
undergo a random walk, ending when the piece touched the aggregate
and stuck on. Iterations of this process made for a very large and
interesting object. Hastings and Levitov represented the addition
process by a conformal map from a circle to a circle with a bump
upon it. When such process is iterated many times, the composition
rule for conformal maps produced a continually growing object.
Since the random walk and conformal function both obey the Laplace
equation, DLA can be represented by choosing the addition point at
random. Hence Hastings and Levitov constructed a stochastic
procedure within the general class considered by Loewner. This
reformulation might well have spurred people on to think about
what would happen if one combined Loewner and stochasticity.

Another major use of analytic function theory for two dimensional
fractals involved noticing that behavior near critical points had
many invariance properties, including invariance under conformal
transformations. According to the work of Polyakov \cite{Polyakov}
and others \cite{BPZ}, the scale invariance characteristic of
behavior near critical points could be quite naturally generalized
to include all non-local shear-free transformations \cite{cardy}
which then implied conformal invariance. A wide variety of
critical problems were analyzed using conformal methods, but with
a few exceptions the work was limited to the behavior of
thermodynamic functions and of point operators like the
magnetization. As noted above, conformal methods were also applied
to such problems as Brownian motion, self-avoiding walks,
percolation and DLA since these were all considered to be somehow
close to critical phenomena. Indeed several of these problems were
shown to be limiting cases of the critical models. But, in recent
years, work on critical phenomena slowed down somewhat because it
was felt that most of the leading problems had been investigated.

\section{Schramm-Loewner Evolution}

However, this view has proven wrong. The area of two-dimensional
(2D) critical phenomena has enjoyed a recent breakthrough. A
radically  new development, called the Schramm-Loewner Evolution
or SLE (also previously called ``stochastic Loewner evolution'')
\cite{schramm,LSW} has provided us new tools and new questions for
criticality in 2D, and also provides us with a new interpretation
of the traditional conformal field theory (CFT) approach.

Examples of systems described by SLE include familiar statistical
models --- Ising, Potts, $O(n)$ model, polymers, --- as well as
``geometric'' critical phenomena like percolation, self-avoiding
random walks, spanning trees and others. The new description
focuses directly on non-local structures that characterize a given
system, be it a boundary of an Ising or percolation cluster, or
loops in the $O(n)$ model. This description uses the fact that all
these non-local objects become random curves at a critical point
in the continuum limit, and these curves may be precisely
characterized by the stochastic dynamics which we shall describe
in a moment.

This is an exciting development in that SLE complements the
earlier approaches to problems in critical phenomena. It appears
that questions that are difficult to pose and/or answer within CFT
are easy and natural in the SLE framework, and vice versa. One of
the challenges of the near future is to extend the overlap between
the two approaches as far as possible and see whether they are
really equivalent.

So what is SLE? It is simply the study of the Loewner equation
with stochastic driving, specifically driving by a $\xi(t)$ which
a Gaussian random variable, obeying the familiar Langevin
equations of Brownian motion,
\begin{equation}
\langle \dot \xi(t) \dot \xi(s) \rangle =
\kappa \delta(t - s).
\label{Brown}
\end{equation}
or equivalently in more integrated form
\begin{equation}
\langle (\xi(t) - \xi(s))^2 \rangle = \kappa |t - s|.
\label{Gauss}
\end{equation}
Here $\kappa$ is a dimensionless constant whose value is very
important determinant of the behavior. It is usual to refer to SLE
at a particular value of $\kappa$ as  SLE$_\kappa$.

It was Schramm's idea \cite{schramm} that one can use Loewner
equation to describe conformally-invariant random curves if one
chooses $\xi(t)$ to be a random function that satisfies certain
conditions. First, $\xi(t)$ must be continuous with probability
one. Secondly, to produce a conformally-invariant curve, the
process $\xi(t)$ has to have independent identically distributed
increments, since one can construct the required map $g_t(z)$ by
iterations of some infinitesimal identically-distributed conformal
maps. Further natural requirement of the invariance of $g_t(z)$
with respect to reflection $x+iy \to -x+iy$ makes the choice of
$\xi(t)$ essentially unique: it can only be a scaled version of
the Brownian motion without drift of Eq. (\ref{Brown}).

\begin{figure}[]
\includegraphics[width=3.4in]{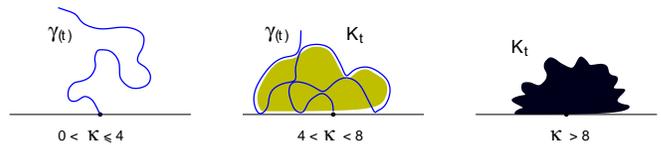}
\caption{Three different traces produced by SLE. This sketch is
taken from Bauer and Bernard ``Conformal field theories of
stochastic Loewner evolutions'', arXiv:hep-th/0210015.}
\label{trace}
\end{figure}

We have already seen that if we chose $\xi(t)$ to be a smooth
real-valued function, the solution $g_t(z)$ of Eq. (\ref{L}) would
give a conformal map from $\mathcal{H}$ cut along a segment traced
out by a simple, non self-intersecting,  curve $\gamma$.
Similarly, in SLE $g$ gives a map from a $\mathcal{D}_t$ in
$\mathcal{H}$  onto $\mathcal{H}$.  This region is $\mathcal{H}$
with some part cut away by the singularities generated in $g$. The
cut out part might be a simple curve which avoids the real line
(for $0< \kappa \leqslant 4$), a self-intersecting one (for
$4<\kappa< 8$), or a filled in region (for $\kappa \geqslant 8$).
For pictures of these possibilities see Figure \ref{trace}. There
are theorems and speculations saying that the trace of the cut-out
singularities or the curves which surround the self-intersection
trace,  or the filled in region gives direct and useful
information about the clusters and other geometrical objects
formed in critical phenomena and other associated scale-invariant
processes.

The theorems apply to percolation, the so-called loop-erased
random walks, and the uniform spanning trees, which are related,
correspondingly, to SLE$_6$, SLE$_2$, and SLE$_8$. They state
essentially that the ensemble of all the appropriate SLE objects
is identical to the conformally-invariant scaling limit of the
critical cluster boundaries for percolation, or of paths --- for
loop-erased random walks and the space-filling curves winding
along the uniform spanning trees. It is also known that if
self-avoiding walks have a conformally-invariant continuum limit,
this limit must be described by SLE$_ {8/3}$.

The corresponding critical phenomena statements (conjectures) are
equally straightforward. They start by connecting each phase
transition problem with a value of $\kappa$. For example, the
standard critical Ising model is said to corresponds to
$\kappa=3$. Various $\kappa$-values are similarly ascribed to the
$O(n)$ model, the $q$-state Potts model (in which case $\sqrt{q} =
- 2\cos(4\pi/\kappa)$), \ldots each at their critical point.  The
the ensemble of traces or boundaries for the SLE process is argued
to be identical with the ensemble of scaling-limit cluster
boundaries for the critical problem. Cardy \cite{cardy-1} has
suggested that the correlation functions of boundary operators in
critical phenomena could be explicitly calculated by using SLE
methods. One of the first SLE-style calculations was due to
Duplantier \cite{Duplantier}, who was interested in the ensemble
of field strengths which would be realized in the neighborhood of
a Potts model or Ising model cluster if that cluster were a
charged conductor. He needed solutions to the Laplace equation in
the exterior of the conductor, but any analytic function
automatically provides such solutions. Hence the mapping functions
$g$ would provide the necessary information. All three of these
calculations, and many more are new results, unexpectedly
available from SLE methods.

We shall not summarize the many important advances made with the
aid of SLE methods, nor speculate about the like advances to come.
Instead we simply say that this is a beautiful area of work, begun
but not completed, where statistical scientists and analysts may
hope to make further important advances.  A bibliography on the
subjects discussed in this paper is included in the appendix.

A review for physics audience on SLE and its relation to discrete
models by Wouter Kager and Bernard Nienhuis is planned for a later
issue of this journal.

\section{Acknowledgements}
This work was primarily supported by the NSF MRSEC Program under
DMR-0213745. At various times the authors have had helpful
interactions with Michel Bauer, Marko Kleine Berkenbusch, Denis
Bernard, John Cardy, Isabelle Claus, Wouter Kager, Bernard
Nienhuis, Ilia Rushkin, and Paul Wiegmann on subjects related to
this note. Figure \ref{sin} was produced by Wouter Kager, who has
kindly given us permission to use it. We thank Michel Bauer and
Denis Bernard for permitting us to use their sketch in figure
\ref{trace}.

\appendix*

\newcounter{items}

\section{References on SLE and related subjects.}

\subsection{History}

The Loewner equation first appeared in the paper

\z K. L\"owner (C. Loewner), Untersuchungen \"uber schlichte
konforme Abbildungen des Einheitskreises, I. MAth. Ann. {\bf 89},
103, 1923.

The equation was proposed as a way of attacking the so called
Bieberbach's conjecture. The conjecture was stated in 1916, and
was finally proven in 1984 by de Branges. The proof relies heavily
on the Loewner's method. There are several books with accounts of
the Loewner's method and the proof of the conjecture. We list
three:

\z L. V. Ahlfors, Conformal invariants: topics in geometric
function theory, New York, McGraw-Hill, 1973.

\z P. L. Duren, Univalent functions, New York, Springer-Verlag,
1983.

\z  Sheng Gong, The Bieberbach conjecture, Providence, R.I.
American Mathematical Society: International Press, 1999.

\subsection{Pure math papers}

\subsubsection{Reviews.}

\z G. F. Lawler, Conformally invariant processes in the plane, a
draft of a book available online at URL
\verb|http://www.math.cornell.edu/~lawler/book.ps|

\z G. F. Lawler, Conformally invariant processes in the plane,
available online at URL
\verb|http://www.math.cornell.edu/~lawler/papers.html|

\z G. F. Lawler, Conformal invariance, universality, and the
dimension of the Brownian frontier, Proceedings of the
International Congress of Mathematicians, Vol. III, 63--72 (Higher
Ed. Press, Beijing, 2002); arXiv: math.PR/0304369.

\z G. Lawler, Introduction to the stochastic Loewner evolution,
available online at URL
\verb|http://www.math.cornell.edu/~lawler/papers.html|

\z O. Schramm, Scaling limits of random processes and the outer
boundary of planar Brownian motion, Current developments in
mathematics, 2000, 233--253 (Int. Press, Somerville, MA, 2001).

\z W. Werner, Critical exponents, \hbox{conformal} in\-vari\-ance
and planar Brownian motion, arXiv: math.PR/0007042.

\z W. Werner, Random planar curves and Schramm-Loewner evolutions,
arXiv: math.PR/0303354.

\z W. Werner, Conformal restriction and related questions, arXiv:
math.PR/0307353.

\subsubsection{Research papers.}

The original paper where the SLE was introduced is

\z O. Schramm, Scaling limits of loop-erased random walks and
uniform spanning trees, Israel J. Math., {\bf 118}, 221--288,
2000; arXiv: math.PR/9904022.

Many more papers followed.

\z R. O. Bauer, Loewner's equation in noncommutative probability,
arXiv: math.PR/0208212.

\z R. O. Bauer, Discrete Loewner evolution, arXiv:
math.PR/0303119.

\z R. O. Bauer, Chordal Loewner families and univalent Cauchy
transforms, arXiv: math.PR/0306130.

\z V. Beffara, On conformally invariant subsets of the planar
Brownian curve, arXiv: math.PR/0105192.

\z V. Beffara, Hausdorff Dimensions for $SLE_6$, arXiv:
math.PR/0204208.

\z V. Beffara, The dimension of the SLE curves, arXiv:
math.PR/0211322.

\z L. Carleson and N. Makarov, Aggregation in the Plane and
Loewner's Equation, Commun. Math. Phys. {\bf 216}, 583 (2001).

\z L. Carleson and N. Makarov, Laplacian path models, J. Anal.
Math. {\bf 87}, 103--150 (2002).

\z J. Dubedat, SLE and triangles, Electron. Comm. Probab. {\bf 8},
28--42 (2003); arXiv: math.PR/0212008.

\z J. Dubedat, Reflected planar Brownian motions, intertwining
relations and crossing probabilities, arXiv: math.PR/0302250.

\z J. Dubedat, $SLE(\kappa,\rho)$ martingales and duality, arXiv:
math.PR/0303128.

\z J. Dubedat, Critical percolation in annuli and $SLE_6$, arXiv:
math.PR/0306056.

\z  R. Friedrich, W. Werner, Conformal fields, restriction
properties, degenerate representations and SLE.  C. R. Math. Acad.
Sci. Paris  {\bf 335}, 947--952 (2002); arXiv: math.PR/0209382.

\z  R. Friedrich, W. Werner, Conformal restriction, highest-weight
representations and SLE, arXiv: math.PR/0301018.

\z P. Kleban and D. Zagier, Crossing probabilities and modular
forms, arXiv: math-ph/0209023.

\z G. F. Lawler, O. Schramm, W. Werner, Values of Brownian
intersection exponents I: Half-plane exponents, Acta Math. {\bf
187}, 237--273 (2001); arXiv: math.PR/9911084.

\z G. F. Lawler, O. Schramm, W. Werner, Values of Brownian
intersection exponents II: Plane exponents, Acta Math. {\bf 187},
275--308 (2001); arXiv: math.PR/0003156.

\z G. F. Lawler, O. Schramm, W. Werner, Values of Brownian
intersection exponents III: Two sided exponents, Ann. Inst. H.
Poincar´e Probab. Statist. {\bf 38}, no 1, 109--123 (2002); arXiv:
math.PR/0005294.

\z G. F. Lawler, O. Schramm, W. Werner, Analyticity of
intersection exponents for planar Brownian motion, Acta Math. {\bf
189}, 179--201 (2002); arXiv: math.PR/0005295.

\z G. F. Lawler, O. Schramm, W. Werner, The dimension of the
planar Brownian frontier is 4/3, Math. Res. Lett. {\bf 8},
401--411 (2001); arXiv: math.PR/0010165

\z G. F. Lawler, O. Schramm, W. Werner, Sharp estimates for
Brownian non-intersection probabilities, in ``In and out of
equilibrium'' (Mambucaba, 2000),  113--131, Progr. Probab. {\bf
51} (Birkhäuser Boston, Boston, MA, 2002); arXiv: math.PR/0101247.

\z G. F. Lawler, O. Schramm, W. Werner, One-arm exponent for 2D
critical percolation, Electr. J. Pobab. {\bf 7}, no 2, 13 pp.
(electronic) (2002); arXiv: math.PR/0108211.

\z G. F. Lawler, O. Schramm, W. Werner, Conformal invariance of
planar loop-erased random walks and uniform spanning trees, Ann.
Probab., to appear; arXiv: math.PR/0112234.

\z G. F. Lawler, O. Schramm, W. Werner, On the scaling limit of
planar self-avoiding walk, arXiv: math.PR/0204277.

\z G. F. Lawler, O. Schramm, W. Werner, Conformal restriction: the
chordal case, J. Amer. Math. Soc., to appear; arXiv:
math.PR/0209343.

\z G. F. Lawler, W. Werner, Intersection exponents for planar
Brownian motion, Ann. Probab. {\bf 27}, 1601-1642.

\z G. F. Lawler, W. Werner, Universality for conformally invariant
intersection exponents, J. European Math. Soc. {\bf 2}, 291--328.

\z G. F. Lawler, W. Werner, The Brownian loop soup, arXiv:
math.PR/0304419.

\z D. Marshall and S. Rohde, The Loewner differential equation and
slit mappings, available online at URL
\verb|http://www.math.washington.edu/~rohde/|

\z S. Rohde, O. Schramm, Basic properties of SLE, arXiv:
math.PR/0106036.

\z O. Schramm, A percolation formula, Electr. Comm. Probab. {\bf
6}, 115--120 (2001); arXiv: math.PR/0107096.

\z S. Smirnov, Critical percolation in the plane: conformal
invariance, Cardy's formula, scaling limits, C.R. Acad. Sci. Paris
S\'er. I Math. {\bf 333}, 239--244 (2001).

\z S. Smirnov, Critical percolation in the plane. I. Conformal
invariance and Cardy's formula. II. Continuum scaling limit,
available online at URL
\verb|http://www.math.kth.se/~stas/papers/|

\z S. Smirnov, W. Werner, Critical exponents for two-dimensional
percolation, Math. Res. Lett. {\bf 8}, 729--744, 2001; arXiv:
math.PR/0109120.

\z W. Werner, Girsanov's transformation for SLE$(\kappa,\rho)$
processes, intersection exponents and hiding exponents, arXiv:
math.PR/0302115.

\z W. Werner, SLEs as boundaries of clusters of Brownian loops,
arXiv: math.PR/0308164.

\subsection{Physics and computational papers}

\subsubsection{Reviews.}

\z J. Cardy, Conformal invariance in percolation, self-avoiding
walks and related problems, arXiv: cond-mat/0209638.

\z B. Duplantier, Conformal fractal geometry and boundary quantum
gravity, arXiv: math-ph/0303034.

\subsubsection{Research papers.}

\z M. Aizenman, B. Duplantier, and A. Aharony, Path-Crossing
Exponents and the External Perimeter in 2D Percolation, Phys. Rev.
Lett. {\bf 83}, 1359 (1999); arXiv: cond-mat/9901018.

\z M. Bauer, D. Bernard, SLE$_k$ growth processes and conformal
field theories, Phys. Lett. B {\bf 543}, 135--138 (2002); arXiv:
math-ph/0206028

\z M. Bauer, D. Bernard, Conformal field theories of stochastic
Loewner evolutions, to appear in Commun. Math. Phys.; arXiv:
hep-th/0210015.

\z M. Bauer, D. Bernard, SLE martingales and the Virasoro algebra,
Phys. Lett. B {\bf 557}, 309--316 (2003); arXiv: hep-th/0301064.

\z M. Bauer, D. Bernard, Conformal transformations and the SLE
partition function martingale, arXiv: math-ph/0305061.

\z J. Cardy, Crossing formulae for critical percolation in an
annulus, J. Phys. A: Math. Gen. {\bf 35}, L565--L572 (2002);
arXiv: math-ph/0208019.

\z J. Cardy, Stochastic Loewner evolution and Dyson's circular
ensembles, J. Phys. A: Math. Gen. {\bf 36}, L379--L386 (2003);
arXiv: math-ph/0301039.

\z B. Duplantier, Two-dimensional copolymers and exact conformal
multifractality, Phys. Rev. Lett. {\bf 82}, 880 (1999); arXiv:
cond-mat/9812439.

\z B. Duplantier, Harmonic measure exponents for two-dimensional
percolation, Phys. Rev. Lett. {\bf 82}, 3940 (1999); arXiv:
cond-mat/9901008.

\z B. Duplantier, Conformally invariant fractals and potential
theory, Phys. Rev. Lett. {\bf 84}, 1363 (2000); arXiv:
cond-mat/9908314.

\z B. Duplantier, Higher conformal multifractality, J. Stat. Phys.
{\bf 110}, 691--738 (2003); arXiv: cond-mat/0207743.

\z B. Duplantier and I. A. Binder, Harmonic measure and winding of
conformally invariant curves, Phys. Rev. Lett. {\bf 89}, 264101
(2002); arXiv: cond-mat/0208045.

\z M. B. Hastings, Exact multifractal spectra for arbitrary
laplacian random walks, Phys. Rev. Lett. {\bf 88}, 055506 (2002);
arXiv: cond-mat/0109304.

\z M. B. Hastings and L. S. Levitov, Laplacian growth as
one-dimensional turbulence, Physica D {\bf 116}, 244 (1998);
arXiv: cond-mat/9607021.

\z T. Kennedy, Monte Carlo tests of SLE Predictions for the 2D
self-avoiding walk, arXiv: math.PR/0112246

\z T. Kennedy, Conformal invariance and stochastic Loewner
evolution predictions for the 2D self-avoiding walk - Monte Carlo
tests, arXiv: math.PR/0207231.

\z M. G. Stepanov and L. S. Levitov,  Laplacian growth with
separately controlled noise and anisotropy, Phys. Rev. E {\bf 63},
061102 (2001); arXiv: cond-mat/0005456.

\end{document}